\documentstyle[multicol,aps,epsf,amstex]{revtex}
\begin{document}
\draft \title{Hydrodynamics of domain growth in nematic liquid crystals} 
\author{G\'eza T\'oth$^1$, Colin Denniston$^2$ and J.M. Yeomans$^1$}
\address{$^1$ Dept. of Physics, Theoretical Physics,
University of Oxford, 1 Keble Road, Oxford OX1 3NP}
\address{$^2$ Dept. of Physics and Astronomy, The Johns Hopkins
  University, Baltimore, MD 21218.}
  
\date{\today} 

\maketitle

\begin{abstract}

We study the growth of aligned domains in nematic liquid crystals. 
Results are obtained solving the Beris-Edwards equations of motion
using the lattice Boltzmann approach. Spatial anisotropy in the domain
growth is shown to be a consequence of the flow induced by the changing
order parameter field (backflow).  The generalization
of the results to the growth of a cylindrical domain, which involves the
dynamics of a defect ring, is discussed.
\end{abstract}

% 83.80.Xz   Liquid crystals: nematic, cholesteric, smectic, discotic, etc 
% 47.11.+j   Computational methods in fluid dynamics
% 61.30.Jf   Defects in liquid crystals   
%{\bf KEY WORDS:} lattice Boltzmann; nematic liquid crystals; complex fluids;
%domain growth; speed anisotropy. 
\begin{multicols}{2}

\section{Introduction}

Liquid crystals \cite{GP93} are an ideal material for the study of topological
defects due to the complex textures they create which are easily
visible to the naked eye.  As topological defects arise in many
situations the observed phenomena in liquid crystals can be
used to test theories in other areas of physics from cosmic strings
\cite{T89} to vortices in superfluid helium \cite{O50}.  Contrary to
the assumption inherent in most previous studies of defect dynamics
in liquid crystals \cite{LM97,D96,IO73}, in a recent Letter
\cite{GDY02} we found that backflow, the coupling
between the order parameter and the velocity fields, has a significant
effect on the motion of defects.  In
particular, the defect speed can depend strongly on the topological
strength in two dimensions and on the sense of rotation of the
director about the core in three dimensions.

These defects were free, in the sense that they were in an unbounded
system.  However, it is much easier to study liquid crystals
experimentally in a confined system.  A very straightforward example
is the geometry used for display 
devices.  In such a display the liquid crystal is sandwiched between
two plates.  As the optical and electrical responses of the liquid
crystal are coupled, one can apply an electric field between the two
plates and directly observe the behaviour. 

The operational state of many such devices, including pi-cells \cite{AT00},
is topologically distinct from its state at zero voltage.  Before the
device can be used, the operational state must be nucleated and grow
to fill the display.  (A typical cross section through a domain wall
separating an operational state and a zero voltage state is shown in
Fig.~\ref{fig_twodefects}(b).) 
These interfaces between topologically distinct
states can behave differently than nematic-isotropic interfaces studied
by other groups \cite{GV99}.
Recent experimental work on pi-cells \cite{AT00} has shown an unusual
anisotropy in the domain growth:  one side of a domain grows faster
than the opposite side.  This would appear to be very similar to the 
anisotropy observed in our simulations of free defects \cite{GDY02}. 
However there are important differences in this system related to the
wall tilt angle, which can dominate the defect-defect interaction
energy, and the driving force of the electric field.  As such, in
order to unambiguously characterize the observed anisotropy
we need to directly study the growth of these domain walls--which
incorporate topological defects--in a confined geometry.

In additional to the technological applications, similar devices
have been proposed as ideal experimental realizations of
two-dimensional (2D) Ising models (in the plane defined by the walls
of the device).  Coarsening of reverse tilt domains in liquid crystal
cells with heterogeneous alignment layers has been shown to be
consistent with predictions of the random-bond Ising model \cite{SS99}
thus providing experimental confirmation of theoretical predictions
for domain growth under conditions of quenched random disorder.

The dynamics of a liquid crystal medium is often modeled 
by using the Ericksen-Leslie-Parodi
equations of motion \cite{GP93,E66}.  These equations describe the state
of the liquid crystal in terms of a director field ${\bf n}$ which is
related to the orientation of the typically long, thin, rod-like
molecules which make up the liquid crystalline material.  
The Ericksen-Leslie-Parodi equations are restricted to an
uniaxial order parameter field of constant magnitude.  Thus
they cannot model the dynamics of topological defects
where in the defect core the magnitude of order has a steep gradient
and the order parameter field is biaxial. (However, they
provide a good description of the bulk away from the defect core.)  

In order to describe the hydrodynamics of topological 
defects correctly, we use the more complex
Beris-Edwards formulation of liquid crystal
hydrodynamics\cite{BE94_BE90}.  The propensity to order, as well as
the direction along which the system orders are conveniently described
by a tensor order parameter ${\bf Q}$ \cite{GP93}. 
The Beris-Edwards equations allow for
variations in the magnitude of the nematic order parameter as well as
biaxiality present in defect cores. They model both
defect dynamics and the coupling between the velocity field and 
the motion of the order parameter.
We use a recent lattice Boltzmann algorithm\cite{DO00} which has
been shown to successfully model the full Beris-Edwards equations.

Our aim in this paper
is to study the growth of a domain of a nematic liquid crystal
at the expense of a second domain with a different director
orientation. Defects form at the walls between domains and their dynamics
is vital in controlling the rate of growth. We find that a spatial
anisotropy in domain growth can result from backflow and discuss
how the wall speed varies with the material parameters of the liquid crystal, 
such as viscosity
and elastic constants, the geometry and the surface properties
of the confining cell and a external electric field.
The Beris-Edwards equations of
motion are presented in Section II, and the results of the domain
growth are described in Section III and Section IV.
In Section V we discuss the relevance of our results to the 
experiments on pi-cells \cite {AT00}.  An outline of the numerical
algorithm is given in the Appendix.
 
\section{The hydrodynamic equations of motion}
\label{2.0}

We summarize the formulation of liquid crystal hydrodynamics described by
Beris and Edwards\cite{BE94_BE90}, extended to include an electric
field and surface potentials. 
Similar models have been examined
by a number of researchers\cite{RT98}.
The continuum equations of motion are written in terms of a tensor order
parameter ${\bf Q}$ which is related to the direction of individual
molecules ${\hat{m}}$ by $Q_{\alpha\beta}= \langle \hat{m}_\alpha
\hat{m}_\beta - {1\over 3} \delta_{\alpha\beta}\rangle$ where the angular
brackets denote a coarse-grained average. (Greek indices will be used to
represent Cartesian components of vectors and tensors and the usual
summation over repeated indices will be assumed.) ${\bf Q}$ is a traceless
symmetric tensor. Its largest eigenvalue, $\frac {2} {3} q$, $0<q<1$,
describes the magnitude of the order. 

We first write down a Landau-de Gennes free energy which describes the
equilibrium properties of the liquid crystal\cite{GP93,DE89}
\begin{equation}
{\cal F}=\int_{V} dV \left\{ f_{bulk}+f_{el}+f_{field}\right\} + 
         \int_{S} dS \left\{ f_{surf}\right\}.
\label{free}
\end{equation}
$f_{bulk}$ describes the bulk free energy
\begin{eqnarray}
f_{bulk}&=&\frac {A_0}{2} (1 - \frac {\gamma} {3}) Q_{\alpha \beta}^2 - 
          \frac {A_0 \gamma}{3} Q_{\alpha \beta}Q_{\beta
          \gamma}Q_{\gamma \alpha} \nonumber\\
&&+ \frac {A_0 \gamma}{4} (Q_{\alpha \beta}^2)^2.
\label{eqBulkFree}
\end{eqnarray}
For $\gamma=2.7$ there is a first-order transition from the isotropic
to the nematic phase.  The minimum of $f_{bulk}$ describes a
uniaxial nematic with an order parameter of the form 
$Q_{\alpha\beta}= q (n_\alpha n_\beta-{1\over 3}
\delta_{\alpha\beta})$ where $q$ is zero in the isotropic phase and
has a finite value in the nematic phase, and ${\bf n}$ is the director field.

$f_{el}$ is the analogue of the Frank elastic free energy density
\begin{eqnarray}
f_{el}=\frac{L_1}{2} (\partial_\alpha Q_{\beta \gamma})^2+
\frac{L_2}{2} (\partial_\alpha Q_{\alpha \gamma})(\partial_\beta Q_{\beta \gamma})+\nonumber\\
\frac{L_3}{2} Q_{\alpha \beta}(\partial_\alpha Q_{\gamma \epsilon})
(\partial_\beta Q_{\gamma \epsilon}).
\label{fFrank}
\end{eqnarray}
This can be easily mapped to give the Frank elastic constants $K_1$,
$K_2$ and $K_3$ \cite{BE94_BE90}.  In particular, the ``one elastic
constant'' approximation, $K_1=K_2=K_3$ corresponds to $L_1>0$ and
$L_2=L_3=0$.

For a uniaxial nematic, the dielectric constant is anisotropic
measured along or perpendicular to the director.  The relation between
the electric displacement ${\bf D}$ and field ${\bf E}$ is of the form
\cite{GP93} 
\begin{equation}
{\bf D}=\epsilon_\perp {\bf E}+(\epsilon_\parallel-\epsilon_\perp)({\bf
  n}\cdot{\bf E}) {\bf n}.
\label{eps}
\end{equation}
More generally, the dependence of the dielectric constant on the
order parameter is described by
\begin{equation}
\epsilon_{\alpha\beta}=\frac{2}{3}\epsilon_a
Q_{\alpha\beta}+\epsilon_m \delta_{\alpha\beta}
\end{equation}
where
\begin{eqnarray}
\epsilon_a &=& \frac{3}{2 q} (\epsilon_\parallel-\epsilon_\perp),\\
\epsilon_m &=& \frac{2}{3}\epsilon_\perp+ \frac{1}{3}\epsilon_\parallel,
\end{eqnarray}
giving results consistent with Eqn.(\ref{eps}) for the uniaxial nematic.
The electric contribution to the thermodynamic potential $f_{field}$  is
\begin{equation}
f_{field}= -\frac{1}{4\pi}\int {\bf D}\cdot d{\bf E}=-\frac{\epsilon_m}{8\pi}
E^2-\frac{\epsilon_a}{12\pi} E_\alpha E_\beta Q_{\alpha\beta}.
\label{f_field}
\end{equation}

The equation of motion for the nematic order parameter is \cite{BE94_BE90}
\begin{equation}
(\partial_t+{\vec u}\cdot{\bf \nabla}){\bf Q}-{\bf S}({\bf W},{\bf
  Q})= \Gamma {\bf H}
\label{Qevolution}
\end{equation}
where $\Gamma$ is a collective rotational diffusion constant.
The first term on the left-hand side of equation (\ref{Qevolution})
is the material derivative describing the usual time dependence of a
quantity advected by a fluid with velocity ${\vec u}$. This is
generalized by a second term 
\begin{eqnarray}
{\bf S}({\bf W},{\bf Q})
&=&(\xi{\bf A}+{\bf \Omega})({\bf Q}+{\bf I}/3)+({\bf Q}+
{\bf I}/3)(\xi{\bf A}-{\bf \Omega})\nonumber\\
& & -2\xi({\bf Q}+{\bf I}/3){\mbox{Tr}}({\bf Q}{\bf W})
\end{eqnarray}
where ${\bf A}=({\bf W}+{\bf W}^T)/2$ and
${\bf \Omega}=({\bf W}-{\bf W}^T)/2$
are the symmetric part and the anti-symmetric part respectively of the
velocity gradient tensor $W_{\alpha\beta}=\partial_\beta u_\alpha$.
${\bf S}({\bf W},{\bf Q})$  appears in the equation of motion because
the order parameter distribution can be both rotated and stretched by
flow gradients. This is a consequence of the rod-like geometry
of the liquid crystal molecules.
$\xi$ is a constant which depends on the molecular
details of a given liquid crystal.

The term on the right-hand side of Eqn.(\ref{Qevolution})
describes the relaxation of the order parameter towards the minimum of
the free energy. The molecular field ${\bf H}$ which provides the driving
motion is related to the derivative of the free energy by
\begin{eqnarray}
{\bf H}&=& -{\delta {\cal F} \over \delta {\bf Q}}+({\bf
    I}/3) Tr{\delta {\cal F} \over \delta {\bf Q}}\nonumber\\ &=&
    {\bf H_{bulk}}+{\bf H_{el}}+{\bf H_{field}}
\label{H(Q)}
\end{eqnarray}  
where 
\begin{eqnarray}
&&{\bf H_{bulk}}  = - A_0 (1 - \frac {\gamma} {3})
 {\bf Q}+ A_0 \gamma \left({\bf Q^2}-
({\bf I}/3)Tr{\bf Q^2}\right) \nonumber\\
&& \qquad -A_0 \gamma {\bf Q}Tr{\bf Q^2},\\
&& ({\bf H_{el}})_{\alpha \beta} = L_1 ({\partial_\gamma}^2 Q_{\alpha \beta}) \nonumber\\
&& \qquad +L_2 \left \{ \frac {1} {2}
(\partial_{\alpha} \partial_{\gamma} Q_{\gamma \beta}  +
 \partial_{\beta}  \partial_{\gamma} Q_{\gamma \alpha}) -
 \frac {1} {3} \delta_{\alpha \beta} 
\partial_{\gamma} \partial_{\epsilon} Q_{\gamma \epsilon}\right \}  \nonumber\\
&& \qquad +\frac {1} {2} L_3 \biggl \{ \partial_\gamma 
( Q_{\gamma \epsilon} \partial_{\epsilon} 
Q_{\alpha \beta})
-(\partial_\alpha Q_{\gamma \epsilon})
(\partial_\beta Q_{\gamma \epsilon}) \nonumber\\
&& \qquad +\frac {1} {3} \delta_{\alpha \beta} (\partial_\eta Q_{\gamma \epsilon})^2
\label{H_elastic}
\biggr \},\\
&& ({\bf H_{field}})_{\alpha \beta} = \frac {\epsilon_a} {12 \pi} 
( E_{\alpha } E_{\beta}  - \frac {\delta_{\alpha \beta}} {3} {E_{\gamma}}^2),
\end{eqnarray}
and $\delta_{\alpha \beta}$ is the Kronecker delta.  We work in a
two-dimensional cross section, assuming that the order parameter does
not change in the perpendicular direction (although the director may
point out of the simulation plane).  In addition,  the symmetry and
zero trace of ${\bf Q}$ is exploited for simplification.  

At the surfaces of the device we assume a pinning potential
\begin{equation}
f_{surf}= \frac {1} {2} \alpha_S (Q_{\alpha \beta}-
                Q_{\alpha \beta}^0)^2.
\label{free_surface}
\end{equation}
We typically take ${\bf Q}^0$ of the form
\begin{equation}
Q_{\alpha \beta}^0=q(n^0_\alpha n^0_\beta-\delta_{\alpha \beta}/3),
\end{equation}
where $q$ is set to the equilibrium bulk value.
This corresponds to specifying a preferred direction ${\bf n}^0$ for
the director at the surface.  There can be other terms in the surface
free energy \cite{Surfaces} and a complete treatment of surface dynamics can
be quite rich \cite{Rey01}.  However, in this paper we will be operating in the
strong pinning limit ($\alpha_S$ large) so that the only effect of the
pinning potential is to furnish an almost fixed value of
$Q_{\alpha\beta}$ at the surface (equal to $Q_0$).  In all cases
studied here, the results are insensitive to the precise value 
of $\alpha_S$, so long as it is large enough to be in the strong pinning limit.

%When computing the functional derivative in Eqn. (\ref{H(Q)}) surface terms
%arise from the integration by parts.  For reference, these are
%\begin{eqnarray}
%&&\delta{\cal F}_d^{surf} = \int_{\partial V} ds\, \delta Q_{\alpha\beta} 
%\left\{L_1 \sigma_\gamma (\partial_\gamma Q_{\alpha\beta}) \right.\nonumber\\
%&& \qquad\qquad\qquad +\frac{1}{2}L_2 [\sigma_\alpha (\partial_\gamma Q_{\gamma\beta})
%+ \sigma_\beta (\partial_\gamma Q_{\gamma\alpha})] \nonumber\\
%&& \qquad\qquad\qquad \left.+L_3 \sigma_\gamma Q_{\gamma\epsilon}(\partial_\epsilon Q_{\alpha\beta})\right\}
%\label{surfterms}
%\end{eqnarray}
%where $\boldsymbol{\sigma}$ is the surface (unit) normal.

The fluid momentum obeys the continuity
\begin{equation}
\partial_t \rho + \partial_{\alpha} \rho u_{\alpha} =0
\label{continuity}
\end{equation}
and the Navier-Stokes equation 
\begin{eqnarray}
&&\rho(\partial_t+ u_\beta \partial_\beta)
u_\alpha = \partial_\beta \tau_{\alpha\beta}+\partial_\beta
\sigma_{\alpha\beta}\nonumber\\
&& \qquad+\eta \partial_\beta((1-3\partial_\rho
P_{0}) \partial_\gamma u_\gamma\delta_{\alpha\beta}+\partial_\alpha
u_\beta+\partial_\beta u_\alpha)
\label{NS}
\end{eqnarray}
where $\rho$ is the fluid density and $\eta=\rho \tau_f/3$ is an isotropic
viscosity (which is controlled by the simulation parameter $\tau_f$
described in the Appendix).  
The form of this equation is not dissimilar to that for a simple
fluid. However the details of the stress tensor reflect the additional
complications of liquid crystal hydrodynamics. 
There is a symmetric contribution
\begin{eqnarray}
\sigma_{\alpha\beta} &=&-P_0 \delta_{\alpha \beta} \nonumber\\
&-&\xi H_{\alpha\gamma}(Q_{\gamma\beta}+{1\over
  3}\delta_{\gamma\beta})-\xi (Q_{\alpha\gamma}+{1\over
  3}\delta_{\alpha\gamma})H_{\gamma\beta}\nonumber\\
&+& 2\xi
(Q_{\alpha\beta}+{1\over 3}\delta_{\alpha\beta})Q_{\gamma\epsilon}
H_{\gamma\epsilon}-\partial_\beta Q_{\gamma\nu} {\delta
{\cal F}\over \delta\partial_\alpha Q_{\gamma\nu}} 
\label{BEstress}
\end{eqnarray}
and an antisymmetric contribution
\begin{equation}
 \tau_{\alpha \beta} = Q_{\alpha \gamma} H_{\gamma \beta} -H_{\alpha
 \gamma}Q_{\gamma \beta} .
\label{as}
\end{equation}
These additional terms can be mapped onto the Ericksen-Leslie
equations to give the Leslie coefficients \cite{DO00}.
The background pressure $P_0$ is constant in the
simulations to a very good approximation ($\pm 1\%$).

The differential equations for order parameter field Eqn. (\ref{Qevolution}) 
and the flow field Eqn. (\ref{NS}) are coupled. The velocity field and
its derivatives appear in the equation of motion for the order
parameter Eqn. (\ref{Qevolution}). 
Unless the flow field is zero, $\vec{u}=0$, the dynamics given by
Eqn. (\ref{Qevolution}) are not relaxational and hydrodynamics play an important role.
Conversely, the order parameter field affects the dynamics of the flow field 
through the stress tensors (\ref{BEstress}) and (\ref{as})
which appear in the Navier-Stokes equation (\ref{NS})
and depend on ${\bf Q}$ and ${\bf H}$. This back-action
of the order parameter field on the flow field is usually
referred to as backflow.  To study these equations we use a lattice
Boltzmann algorithm summarized in the Appendix.  Other than when
explicitly stated, the simulation parameters are those listed in
\cite{SYM1}.

\section{Domain growth}

\begin{figure}
\narrowtext 
\centerline{\epsfxsize=2.6in 
\epsffile{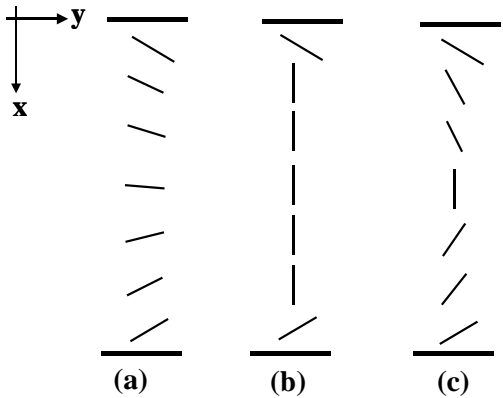}}
\vskip 0.2true cm
\caption{The possible alignment of directors for a tilt angle
  $-\theta_p$ on the top surface and $+\theta_p$ on the bottom surface
  ($\theta_p < 45 \deg$; the surface tilt angle is measured
  with respect to the horizontal axis): 
  (a) Director configuration when the field is switched off and the
  system had time to relax to its global minimum (H, or horizontal state); 
  (b) the field is switched on at a fairly high voltage $ \sim 6 V$
  (V, or vertical state);
  (c) the field is at a voltage $ \sim 2 V$ or lower.  The system
  may remain in the metastable state (c) for some time even at zero
  voltage.  Periodic boundary conditions apply in the horizontal ($y$)
  direction.} 
\label{fig_surf2}
\end{figure}

We consider a liquid crystal confined between two planes a distance $L_x$
apart.  The director field may take topologically distinct states depending on
the boundary conditions and applied voltage.   In the simulations we
set the boundary condition (${\bf Q}^0$ in Eq. (\ref{free_surface}))
so as to give a tilt angle $-\theta_p$ between the director and the 
$y$ axis at $x=0$ and $+\theta_p$ at $x=L_x$. 
At zero applied voltage these conditions result in a global minimum
free energy state with a splayed director configuration, or horizontal
(H) state as shown in Figure~\ref {fig_surf2}(a).  At high voltages,
typically on the order of $~6 V$, the H state is no longer the global
minimum, and a bend configuration (vertical state) is obtained like
the one shown in Figure~\ref {fig_surf2}(b).  At intermediate
voltages, the vertical (V) state is more relaxed as shown in
Figure~\ref {fig_surf2}(c).  

As the H and V states are topologically distinct, the transition from V to H
requires nucleation of H domains and the generation of defects.
 The problem we will
investigate is the growth (or shrinking) of the H state within the V
state.   In
particular, we are interested in how hydrodynamics affects the speed
of the domain walls.  This is partly motivated by the observation in
Ref. \cite{AT00} that the domain growth in a liquid crystal device can
be anisotropic and the speculation that this may be due to
hydrodynamics.  

We have previously observed that the velocity of
defects in unbound systems can be affected by hydrodynamics.
In particular the defect speed can depend strongly on the topological
strength in two dimensions and on the sense of rotation of the
director about the core in three dimensions \cite{GDY02}. 
The crucial difference between the domain growth problem and the motion of
free defects is that in the latter case each defect moves due the
director field of the other. In the domain growth problem
the defects are not interacting but are dragged by the free-energy-driven 
movement of the domain walls. The free defects are accelerated as they
approach each other while in the domain growth problem the defects
move with a constant speed. 
Due to these differences and the additional geometrical parameters, 
a separate analysis is needed for the confined system which is
also easier to realize experimentally and provides a better control
of the parameters influencing the defect speed.   

In order to study the role of hydrodynamics in the system we will
examine the factors affecting 
the domain growth so that we can clearly identify what causes the
wall speed anisotropy.  The key parameters are the surface director tilt
$\theta_p$, the sample thickness $L_x$ and material parameters such as
coefficients in the bulk free energy  (\ref{free}): $A_0$, $\gamma$,
and elastic constants $L_1$, $L_2$ and $L_3$.  In addition, the rotational
diffusion constant $\Gamma$, which appears in the dynamical equation
(\ref{Qevolution}) for the order parameter gives an overall (inverse)
time scale and is related to the Leslie-Ericksen viscosities \cite{DO00}.  

\begin{figure}
\centerline{\epsfxsize=3.0in \epsffile{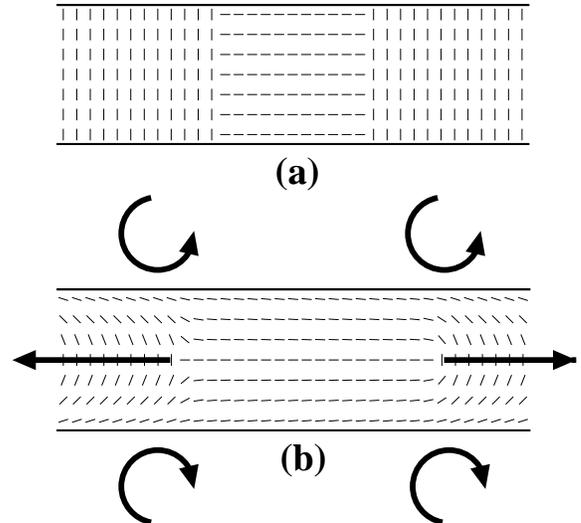}}
%\epsfclipon \epsffile[160 250 700 820]{defectdyn2001_2defects.eps}}
%\includegraphics[width=3.5in]{defectdyn2001_2defects.eps}
\caption{(a) The initial director configuration is a horizontal domain
  (H state) in an otherwise vertically aligned system (V state). 
  (b) As the system begins to relax, two defects are formed at 
  the boundary of the horizontal and vertical domains. The left (right)
  defect has a topological strength $s= -\frac {1} {2}$ ($s= +\frac {1} {2}$). 
  The curved arrows indicate the direction of the vortices induced by
  the reorientation of the director during the growth of the
  horizontal domain.  The straight arrows point into the direction of
  defect motion.  Note that there are periodic boundary conditions in
  the horizontal ($y$) direction.}
\label{fig_twodefects}
\end{figure}

For simplicity, we will first study the undriven case of an H domain
growing at the expense of a V state.  We will later examine growth under the
influence of an electric field.
The initial configuration, depicted in Fig. \ref{fig_twodefects}(a), is a
horizontal (i.e., along the $y$-direction) domain in an otherwise
vertically aligned state. This models a time shortly after the electric
field has been switched off when small but macroscopic domains have formed
in the device.  
As the simulation proceeds, the director configuration relaxes rapidly to
that shown in Fig. \ref{fig_twodefects}(b). 
During the relaxation defects are formed at the center of each domain
wall with strengths $+\frac {1} {2}$ and $-\frac {1} {2}$, respectively. 
Once the two defects have formed the vertical domain
begins to grow and the $+\frac {1} {2}$ and $-\frac {1} {2}$ 
defects move in opposite directions.  

Our simulations correspond to a two-dimensional cross section of the 
two line defects, assuming that the order parameter does not change 
in the perpendicular direction (although the director may point out of the
simulation plane). The two defects are topologically distinct 
only in two dimensions, but even in three dimensions they are usually
separated by an energy barrier.

A particular advantage of the simulations is that
the backflow can easily be switched off by setting 
$\sigma_{\alpha \beta}=-P_0 \delta_{\alpha \beta}$ and 
$\tau_{\alpha \beta}$ to zero. (Compare this to (\ref{BEstress}) and (\ref{as}).) 
Since there is no flow imposed, there is
a zero velocity field throughout the whole simulation. 
The dynamical equation in this case can be obtained
from (\ref{Qevolution}) by setting ${\vec u}$ to zero.
It corresponds to the purely relaxational Ginzburg-Landau model\cite{B94}
\begin{equation}
\partial_t {\bf Q} = \Gamma {\bf H}
\label{Qevolution_noflow}
\end{equation}
where the molecular field ${\bf H}$ is given by (\ref{H(Q)}).
Comparing the dynamics obtained from the Ginzburg-Landau model
and the full hydrodynamic equations,
the effect of the backflow can be unambiguously identified.

The Ginzburg-Landau equation (\ref{Qevolution_noflow}) with a single elastic constant is invariant
under a local coordinate transformation mirroring the director on the $x$
axis. This corresponds to the transformation
\begin{equation}
Q_{xy}\rightarrow -Q_{xy},\;\;\;\;\;\;\;\;\;\;\;\;
Q_{yx} \rightarrow -Q_{yx}.
\label{mapping}
\end{equation}
The order parameter fields of the two {\it moving} defects 
with topological 
charges $s=\pm \frac {1} {2}$ shown in Fig. \ref{fig_twodefects}(b) 
(even including the deformation due to the boundaries) 
transform into each other. Thus approaches based on 
a simple Ginzburg-Landau equation predict
that as the defects move they follow symmetric dynamical
trajectories.

We can construct a simple model for the domain motion in the
absence of hydrodynamic flow.   In the bulk regions (away from the
domain walls), $f_{bulk}$ is minimized with an uniaxial 
order parameter of the
form $Q_{\alpha\beta}= q (n_\alpha n_\beta-{1\over 3}
\delta_{\alpha\beta})$.  We can then restrict our attention to the
elastic free energy $f_{el}$.   With this form of the order parameter
and if $L_2=L_3=0$, the elastic free energy density has the form $f_{el}= L_1
q^2 (\nabla \theta)^2$ if the director remains in the plane so that
${\bf n}=(\cos\theta, \sin \theta,0)$.  The minima
in the bulk regions (away from the domain walls) correspond to the director
angle changing linearly along the $x$ coordinate from $-\theta_p$ to
$+\theta_p$ in the H domain, and from $-\theta_p$ to
$(+\theta_p-180\deg)$ in the V domain.  The difference of the free
energy densities of the two domains can then be written as 
\begin{equation}
\Delta f=f_{V}-f_{H}=\frac {4 \pi L_1 q^2 } {L_x^2} (\frac {\pi} {4}
-\theta_p).
\label{eq_deltaf}
\end{equation}
For  $\theta_p<45\deg$, the horizontal domain grows because this
decreases the free energy of the system.  For $\theta_p>45\deg$ the
horizontal domain should begin to shrink, and at $\theta_p=45\deg$ 
the two domains have the same free energy and the defects
should stop moving.

If the H domain grows by a length of $\Delta L_y$
then the free energy of the system decreases by $\Delta f \times L_x
\times \Delta L_y$.  Simple relaxational arguments then suggest
that the speed of domain growth can be described by the formula 
\begin{equation}
v=\frac {1} {\eta_e} \times \Delta f \times L_x,
\label{eq_v_eta_f}
\end{equation}
where $\eta_e$ is an effective viscosity. 

\begin{figure}
\narrowtext 
\centerline{\epsfxsize=\columnwidth \epsfxsize=3.2in \epsffile{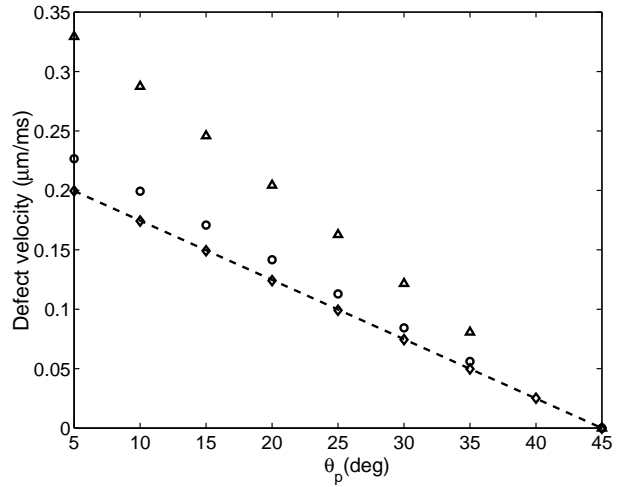} }
\vskip 0.2true cm
\caption{The velocity of the two defects as a function 
of surface tilt if backflow is ignored (diamonds)
or included.  Note that if backflow is not included  
then the two defects move with the same speed, which is well described
by the dashed line based on Eqns. (\ref{eq_deltaf}) and 
(\ref{eq_v_eta_f})
Hydrodynamics accelerates the $s= +\frac {1} {2}$ 
defect (triangles) substantially, while it affects
the $s= -\frac {1} {2}$ defect (circles) much less.
The speed anisotropy $\alpha$ is $36\%$.}
\label{fig_defectspeed_tilt}
\end{figure}

{\bf Surface tilt:}
We first investigate the effect of the surface tilt $\theta_p$ on the defect
speed.  Equations (\ref{eq_deltaf}) and (\ref{eq_v_eta_f}) suggest
that as $\theta_p$ increases, the free energy difference decreases,
and the defects should move more slowly. 
The defect speed $v$ is plotted as a function of surface tilt $\theta_p$
in Fig. \ref{fig_defectspeed_tilt}.
Consider first the diamonds. 
These correspond to the case with backflow switched off. For this case
both defects move at the same speed (but in opposite directions).
Notice that the defect velocity is proportional to
$(\theta_p-45\deg)$. From (\ref{eq_deltaf}) and 
(\ref{eq_v_eta_f}) this leads to the conclusion that the effective
viscosity $\eta_e$ is independent of the tilt angle and its value is
found to be $0.138$ Pa s for the parameters of the simulation.

{\bf Back-flow:} The triangles and circles in
Fig. \ref{fig_defectspeed_tilt} show the 
velocity of the defects when backflow is included in the model. 
The $s=+ \frac {1} {2}$ defect is considerably speeded up, whereas
the $s=- \frac {1} {2}$ defect is only slightly accelerated.
The defect speed remains proportional to $(\theta_p-45\deg)$ 
within a $2\%$ error.  The effective viscosities are
$\eta_{+1/2}=0.083$ Pa s $< \eta_{-1/2}=0.123$ Pa s.  These values are
comparable to the rotational viscosity $\gamma_1=0.08$ Pa s\cite{AT00}.
The speed anisotropy defined as
\begin{equation}
\alpha= \frac {\Delta v} {\overline v}=
\frac {v_{s=+1/2}-v_{s=-1/2}} {(v_{s=+1/2}+v_{s=-1/2})/2}
\end{equation}
is independent of $\theta_p$ and its value is $\alpha=36\%$.

The order parameter field affects the flow field through the symmetric
and antisymmetric stress tensors, (\ref{BEstress}) and (\ref{as}).
The total nonviscous stress (i.e. the combination of all the stress
terms not related to the velocity gradient tensor) is the sum of three
terms
\begin{eqnarray}
{\boldsymbol \sigma} + {\boldsymbol \tau}=
\boldsymbol {\sigma}_{i}+\boldsymbol{\sigma}_{H}+\boldsymbol{\sigma}_{d}.
\label{totalstress}
\end{eqnarray}
Here $\sigma_{i,\alpha \beta}=-P_0 \delta_{\alpha \beta}$ is the
stress due to the isotropic pressure. 
$\sigma_{d,\alpha \beta}=-\partial_\beta Q_{\gamma\nu} {\delta
{\cal F}\over \delta\partial_\alpha Q_{\gamma\nu}}$ is the deformation stress.
For $L_2=L_3=0$ the deformation stress is  
$\sigma_{d,\alpha \beta}=-L_1 Tr (\partial_{\alpha} {\bf Q} \partial_{\beta} {\bf Q}).$
The rest of the terms in (\ref{BEstress}) and (\ref{as}) give what we
will refer to as the molecular field stress,
$\boldsymbol{\sigma}_H$, which is a function of ${\bf H}$ and ${\bf Q}$.
$\boldsymbol{\sigma}_d$ and the diagonal $\boldsymbol{\sigma}_i$ 
do not change under the transformation (\ref{mapping}) that transforms
the defects of topological charge $\pm 1/2$ into each other.
Conversely, the off-diagonal elements of $\boldsymbol{\sigma}_H$
have their sign inverted.  
Thus the stress fields and the resultant backflow are different for
the two defects. 

The stress field $\boldsymbol{\sigma}_d$ is related to the deformation
free energy density, which is the same for both defects. 
It induces vortices similar to those around a solid cylinder moving
in a viscous liquid. The flow points in the direction of defect
motion at the defect core.  The contribution ${\boldsymbol \sigma}_H$ 
describes the stress due to the the reorientation of the director. The
reorientation is the strongest around the core while molecules
near the surfaces reorient much less. The director reorientation
induces vortices around the two defects as shown in
Fig. \ref{fig_twodefects}(b).  The direction of these vortices is
determined by the gradient of the director angle taken moving around
the defect in the positive direction. It is positive (negative) for
the $+\frac {1} {2}$ ($-\frac {1} {2}$) defect. 

\begin{figure}
\narrowtext 
\centerline{\epsfxsize=1.6in 
\epsffile{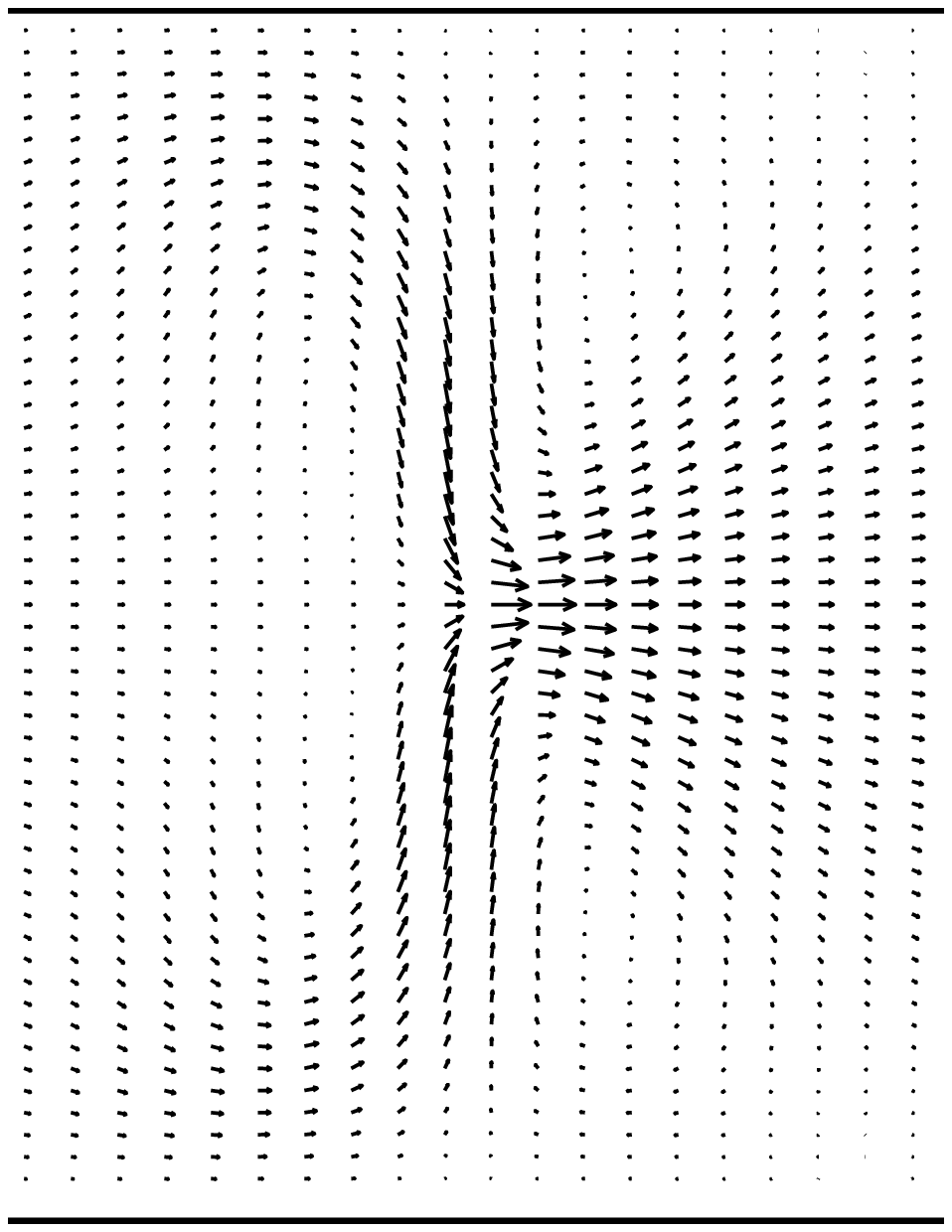} 
\hskip 0.5cm
\epsfxsize=1.6in 
\epsffile{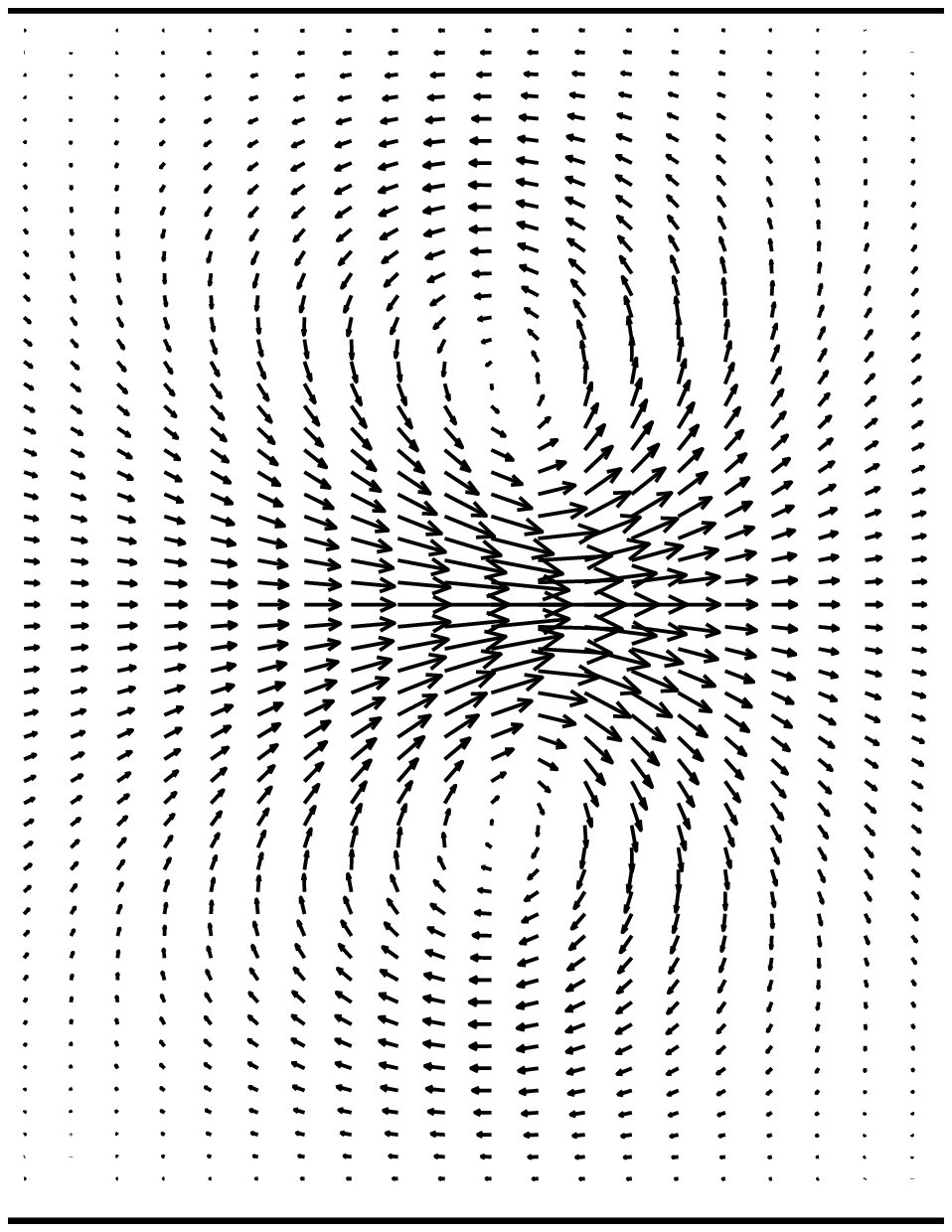}}
\centerline{(a) \hskip 1.6in (b)}
\vskip  0.2true cm
\caption{Velocity field corresponding to the (a) $s= -\frac {1} {2}$ 
and (b) $s= +\frac {1} {2}$ defects shown in Fig. \ref {fig_twodefects}(b).   
There is a strong vortex pair around the $s= +\frac {1} {2}$ defect
which, at the defect core, points in the direction of defect movement. 
The flow at the core of the $s= -\frac {1} {2}$ defect
is weaker and points opposite to the direction of defect propagation.}
\label{fig_velocityfield}
\end{figure}

The two contributions to the backflow reinforce each other for the
$s=+\frac{1}{2}$ defect but partially cancel for the $s=-\frac{1}{2}$ defect.
The resulting flow fields can be seen in
Fig. \ref{fig_velocityfield}. The flow is stronger around the $s= +
\frac {1} {2}$ defect.  Around the $s= - \frac {1} {2}$ defect the
flow is much weaker and the flow field points opposite to the defect
propagation at the core. However, even in this case backflow accelerates 
the relaxational dynamics.

\begin{figure}
\narrowtext 
\centerline{\epsfxsize=3.2in
  \epsffile{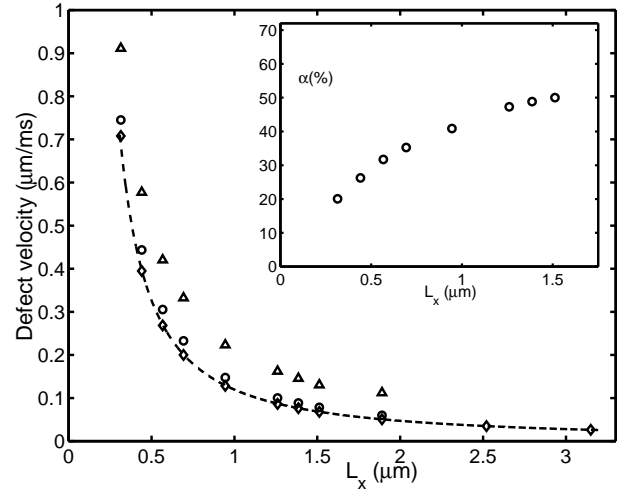} }
\centerline{(a)}
\vskip  0.5true cm
\centerline{\epsfxsize=3.2in
  \epsffile{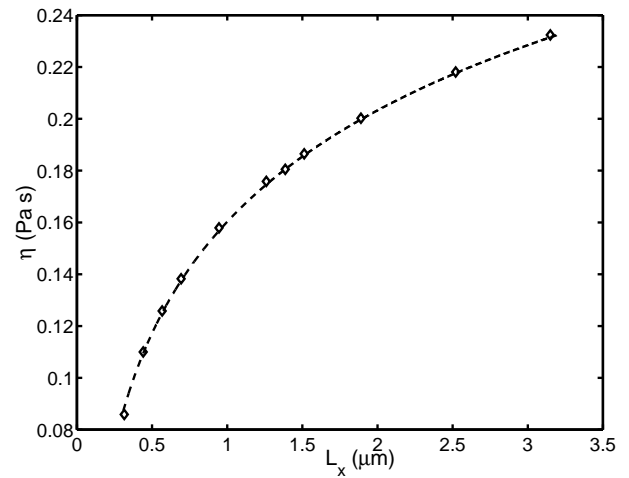} }
\centerline{(b)}
\vskip 0.2true cm
\caption{(a) Speed of the $+\frac {1} {2}$ (triangles) and 
  $-\frac {1} {2}$ (circles) defects as a function of the thickness of
  the sample. The diamonds correspond to the case without hydrodynamics. 
  The inset shows the relative speed anisotropy as a function of 
  sample thickness.  (b) Effective viscosity $\eta_e$ as a function of
  sample thickness for the case without hydrodynamics. 
  The dashed lines in both figures correspond to the fit
  to the theoretical results discussed in the text.}
\label{fig_defectspeed_thickness}
\end{figure}

{\bf Sample dimensions:}  As the sample becomes wider the speed of
the defect propagation decreases \cite{refLx} as can be seen in 
Fig.~\ref{fig_defectspeed_thickness}(a).
The dependence of the defect velocity on $L_x$ follows from
Eqns. (\ref{eq_deltaf}) and (\ref{eq_v_eta_f}) which give
\begin{equation}
v \propto {1 \over \eta_e L_x}.
\label{vLx}
\end{equation}
The effective viscosity can be calculated from 
$\eta_e=1/(2\pi s L_1 \Gamma)\int (\nabla \theta_q)^2 d{\bf r}$, where
$\theta_q$ is the field due just to the defect itself \cite{D96,IO73} and
the integral is over the volume of the system.  Due to the confining
geometry, one expects the integral to be dominated by the near-field
contribution (i.e. the field near the defect core) which is the same
for both a static or moving defect \cite{D96}.  The gradient of
$\theta$ caused by a static defect drops off as $1/r$ and hence one
expects the effective viscosity to go like $\log(L_x/L_{x0})$
\cite{IO73}. 

This dependence is observed in
Fig.~\ref{fig_defectspeed_thickness}(b).  The fit is 
$\eta_e=0.62$ Pa s $\times \log (L_x/L_{x0})$
where $L_{x0}=0.076\mu$m is comparable to the defect core
diameter.  
The dashed line in Fig.~\ref{fig_defectspeed_thickness}(a)
shows the fit to Eq.~(\ref{vLx}) including the $L_x$ dependence of the
viscosity. 

When backflow is considered, the relative speed anisotropy increases
with $L_x$ and saturates at about 60\% as shown in the inset of
Fig.~\ref {fig_defectspeed_thickness}(a).  The increase is probably
due to the increasingly larger regions around the core involved in
director reorientation.  This leads to stronger vortices and hence
stronger hydrodynamic effects.

\begin{figure}
\narrowtext 
\centerline{\epsfxsize=\columnwidth \epsffile{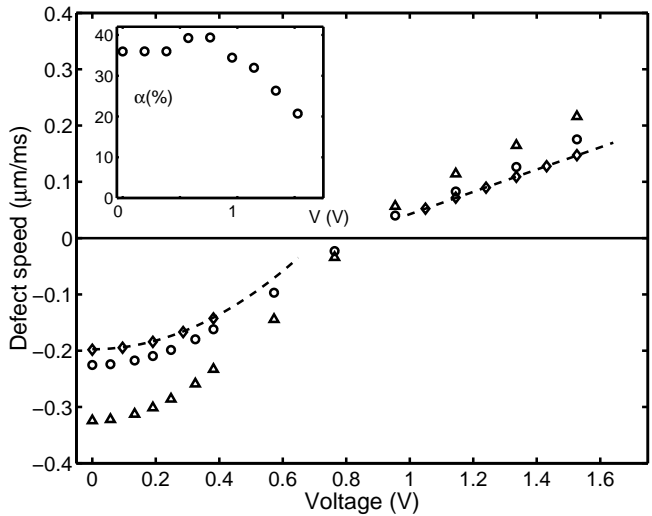} }
\vskip 0.2true cm
\caption{Speed of the $+\frac {1} {2}$ (triangles) and $-\frac {1}
  {2}$ (circles) defects as a function of the external field $V$.
  Without hydrodynamics the defects move at the same speed (diamonds). 
The inset shows the relative speed anisotropy as a function of $V$.}
\label{fig_defectspeed_V}
\end{figure}

{\bf Electric Field:}
When an external electric field is applied, it changes the free energy
densities of the H and V domains.  Thus it also influences the speed
of the domain growth. Fig. \ref{fig_defectspeed_V} shows the speed
of the two defects as the function of applied voltage.
For low voltages the free energy difference between the H and V
domains can be estimated.  If we assume that the orientation of the H
and V states are unchanged from the zero voltage case (i.e. the
director angle remains a linear function of $x$, as used in
Eq. (\ref{eq_deltaf})), then substituting this into Equation
(\ref{f_field}) and integrating over space gives
\begin{eqnarray}
\Delta F_{field} &=& \frac {\epsilon_a q} {48 \pi L_x}  \left(\frac
 {1} {\theta_p} -\frac {1} {\pi/2-\theta_p} \right) V^2 L_y \sin 2 
 \theta_p,
 \label{DF_Field}
\end{eqnarray}
for a sample of length $L_y$.
The total free energy difference between the domains is the sum of the
elastic and the field contributions, Eqn. (\ref{eq_deltaf}) and
Eqn. (\ref{DF_Field}), respectively.  Substituting this into
Eqn. (\ref{eq_v_eta_f}) results in a parabolic dependence of the
velocity on $V$ for low voltages as shown in 
Fig. \ref{fig_defectspeed_V}.
At $V_{limit} \sim 0.9V$ domain growth is reversed since the effect of
the surfaces is balanced by the influence of the electric field.  

\begin{figure}
\narrowtext 
\centerline{\epsfxsize=\columnwidth \epsffile{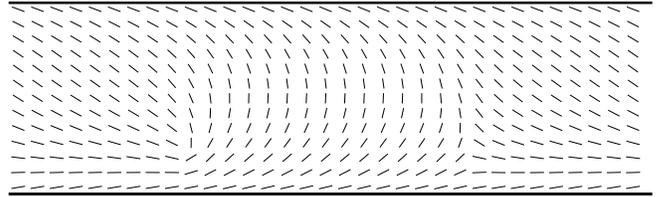}}
\vskip 0.2true cm
\caption{A V domain growing in an asymmetric ${\rm H}_{\rm A}$ environment. 
At high fields the horizontal domain is deformed moving the defects
towards one of the boundaries.}
\label{fig_highV}
\end{figure}

At high voltages ($V\sim 0.6V$), the H domain is 
replaced by ${\rm H}_{\rm A}$, an asymmetric horizontal domain shown
in Fig. \ref{fig_highV}. It has a lower free energy than the H domain 
due to the more favorable alignment with the electric field. 
As a result of the deformation of the ${\rm H}_{\rm A}$ domain, the
defects move towards the surfaces.   
For $V > 1V$ the molecules in the bulk are almost completely aligned
with the vertical field. The horizontal region is then confined to a thin
layer near the surfaces.  For this type of configuration,
the free energy difference between the domains is proportional to the
voltage\cite{AT00}.  This gives the linear slope of the curve in
Fig.~\ref{fig_defectspeed_V} seen for the higher voltages.

For low voltages the speed anisotropy is $36\%$ and independent of the
voltage.  As shown in the inset of Fig.~\ref{fig_defectspeed_V}, 
at high voltages ($V>1V$) the speed anisotropy
decreases. 
The effect on the anisotropy can be explained
by the relative weight of the relaxational dynamics and the backflow.
In the $V=1V$ to $1.6V$ range the relaxational dynamics are substantially speeded up
by the increasing voltage due to the electric field 
contribution (\ref{f_field}) in the free energy. 
The backflow is induced by the 
the stress fields given in
Eq. (\ref{BEstress}) and Eq. (\ref{as}).
These stress fields do not depend directly on the electric field,
only on the order parameter field, which changes only slightly
in this voltage range.  
Therefore the stress fields 
do not increase with the increasing voltage, and the hydrodynamics
is dominated by the relaxational dynamics at high fields.
(In comparison, for the experiment presented in \cite{AT00} the domain
wall speed was  $\sim 0.1 \mu m/ms$ and the anisotropy also decreased
with increasing voltage.)

\section{Other Control Parameters}

The equations governing the dynamics of liquid crystals covered in
Section~\ref{2.0} contain a large set of parameters.  In
this section we explore some of this parameter space.  In particular,
we examine the case of multiple elastic constants, important for
comparing to any real liquid crystal.  In addition we look at the
influence of the different viscosities and free energy parameters on the
balance of the relaxational dynamics and the backflow. 
We also examine the case of asymmetric and inhomogeneous surface tilts
since they give new insights 
about the underlying symmetries of the system and
are important 
for practical devices with non-trivial surface anchoring.

{\bf Elastic constants:}
Real liquid crystals have multiple elastic constants.  We now consider
the effect of non-zero elastic constants $L_2$ and $L_3$.
If $L_2\neq 0$ in the expression of the molecular field
Eqn. (\ref{H_elastic}), the dynamical equation in the absence of backflow
Eq. (\ref{Qevolution_noflow}) loses its invariance under the
transformation (\ref{mapping}).  However, the speed anisotropy is
very small. The reason for this is that the relaxational dynamics are still
invariant under the mirroring transformation for a uniaxial order parameter 
with a constant magnitude \cite{L3=0}. In our setup these conditions
hold except for close to the defect cores. 

\begin{figure}
\narrowtext 
\centerline{\epsfxsize=\columnwidth \epsffile{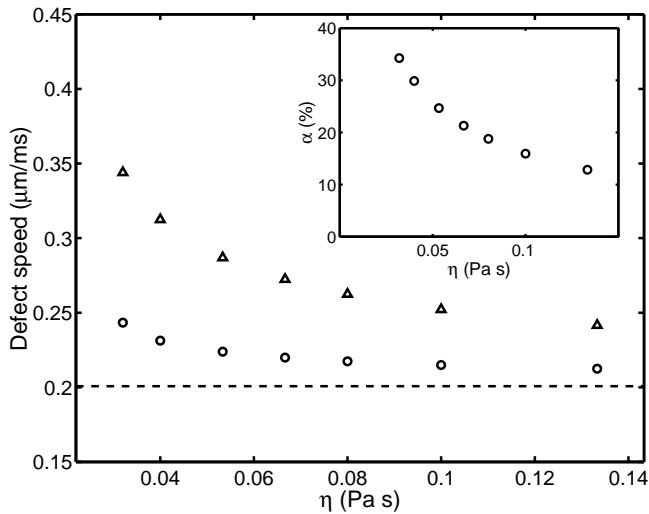} }
\vskip 0.2true cm
\caption{Speed of the $+\frac {1} {2}$ (triangles) and $-\frac {1} {2}$ (circles) defects
as a function of the viscosity
$\eta=\frac {\rho \tau_f} {3}$. 
The dashed line corresponds to the velocity without backflow.
The inset shows the relative speed anisotropy as a function of
$\eta$. }
\label{fig_defectspeed_tau1}
\end{figure}

A larger anisotropy in speed is obtained if $L_3 \neq 0$.  If $L_3>0$
($L_3<0$) then the $s= + \frac {1} {2}$ ($s= - \frac {1} {2}$) defect is
faster.  For $L_1=$8.73pN and $L_3=$15.87pN, we measure a speed
anisotropy of $\alpha=3\%$, in a system without hydrodynamics.  
This anisotropy due to the unequal elastic constants increases with
increasing sample thickness.  For $L_x=1.25 \mu$m (our benchmark
system has $L_x=0.7 \mu$m \cite{SYM1}) the anisotropy 
due to non-zero $L_3$ is $\alpha=6\%$. This may be due to the
fact that, for free defects, the elastic anisotropy causes significant deviations
from the case of isotropic elasticity in the order parameter field away from the axis determined
by the two defect cores\cite{GP93}.  In the thinner sample, the
surfaces "cut off" this part of the field, decreasing the anisotropy. 

\begin{figure}
\narrowtext 
\centerline{\epsfxsize=\columnwidth \epsffile{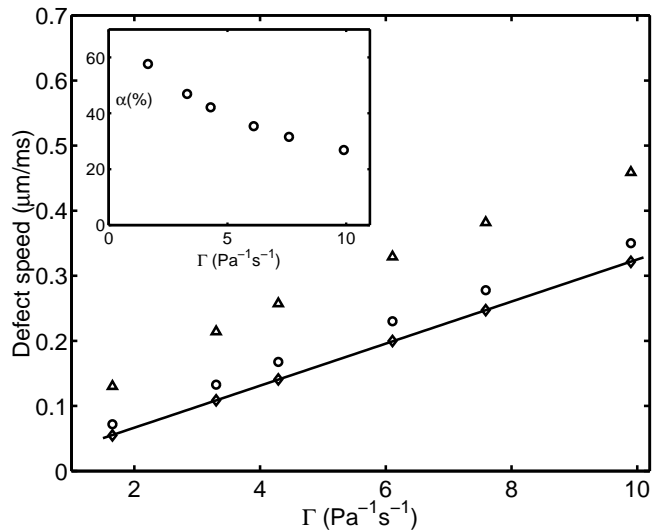} }
\vskip 0.2true cm
\caption{Speed of the $+\frac {1} {2}$ (triangles) and $-\frac {1} {2}$ (circles) defects
as a function of $\Gamma$.  
The diamonds correspond to the case without
hydrodynamics when $v \propto \Gamma$.
The inset shows the relative speed anisotropy as a function of $\Gamma$.}
\label{fig_defectspeed_Gamma}
\end{figure}

If the surface
tilt is close to vertical, and the horizontal domain is shrinking, then
the $L_3$ dependence of the speed anisotropy is the opposite.  If $L_3<0$
($L_3>0$) then the $s= + \frac {1} {2}$ ($s= - \frac {1} {2}$) defect is
faster. Since in the two cases (growing vs. shrinking domain)
the order parameter fields near the axis of the two cores
are the same, this should also be attributed to the differences in the 
order parameter fields far from the defect cores which results from
the different tilts.  

{\bf Viscosities and diffusion:}
Consider now the effect of the parameters governing the time scales in
the equation of motion for the domain growth. 
$\tau_f$ is proportional to the viscosity in the Navier-Stokes
equation (\ref{NS}). Increasing $\tau_f$ increases the viscosity,
slows the defects, and decreases their velocity anisotropy as shown in
Fig. \ref {fig_defectspeed_tau1}.
The velocity tends to that measured without backflow, as represented by
a dashed line in the figure. This is as expected since backflow
will be suppressed by a large viscosity.

Increasing $\Gamma$ which appears in Eq. (\ref{Qevolution}) increases
the velocity of both defects, but decreases the relative speed
anisotropy as shown Fig.~\ref {fig_defectspeed_Gamma}.  The defects
move faster, due to the fact that $\Gamma$ governs the speed of the
relaxational dynamics.  
Since the stress fields in
Eq. (\ref{BEstress}) and Eq. (\ref{as}) do not depend on $\Gamma$,
they do not increase.  As a result, as $\Gamma$ increases, the
relaxational dynamics speed up, but the backflow does not change
as significantly, and as a result the anisotropy decreases.

{\bf Free energy:} $\gamma$ is the parameter in the free energy
(\ref{eqBulkFree}) which controls the magnitude of order in the bulk
of the domains.  (The isotropic-nematic phase transition is at $\gamma=2.7$.)
When $\gamma$ decreases, the defect core gets bigger and
this results in a smaller effective viscosity \cite{D96}.
Thus the defects move faster under the relaxational dynamics. 
The decrease in the magnitude of order results in a smaller
backflow due to the reorientation and $\alpha$ decreases. 

\begin{figure}
\narrowtext 
\centerline {\epsfxsize=\columnwidth
  \epsffile{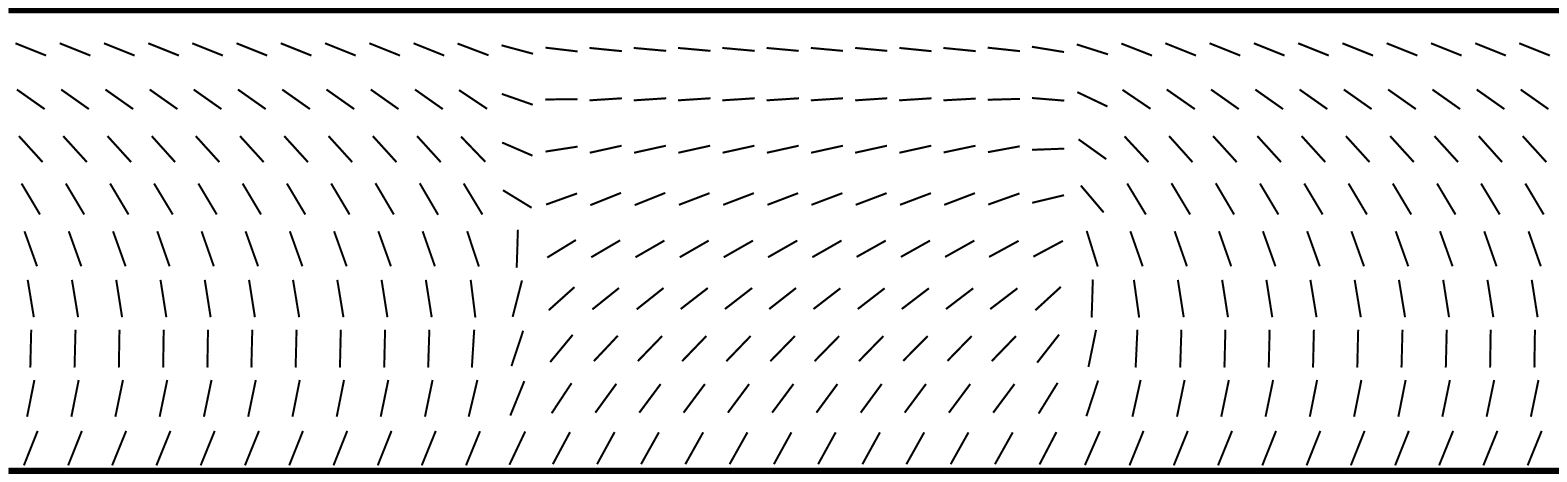} }
\vskip 0.0true cm
\centerline{(a)}
\vskip  0.5true cm
\centerline  {\epsfxsize=1.6in
  \epsffile{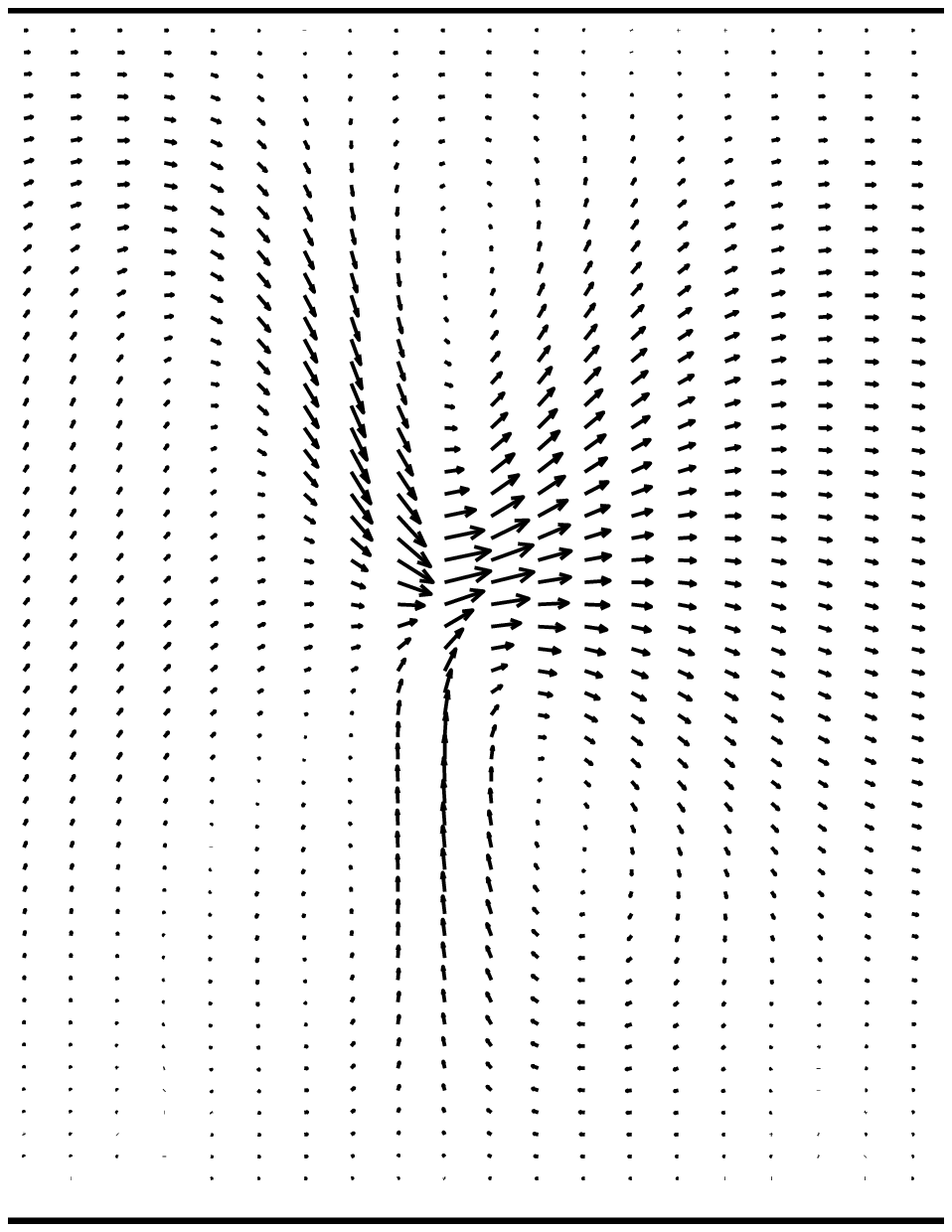}  
  \hskip 0.5cm 
  \epsfxsize=1.6in \epsffile{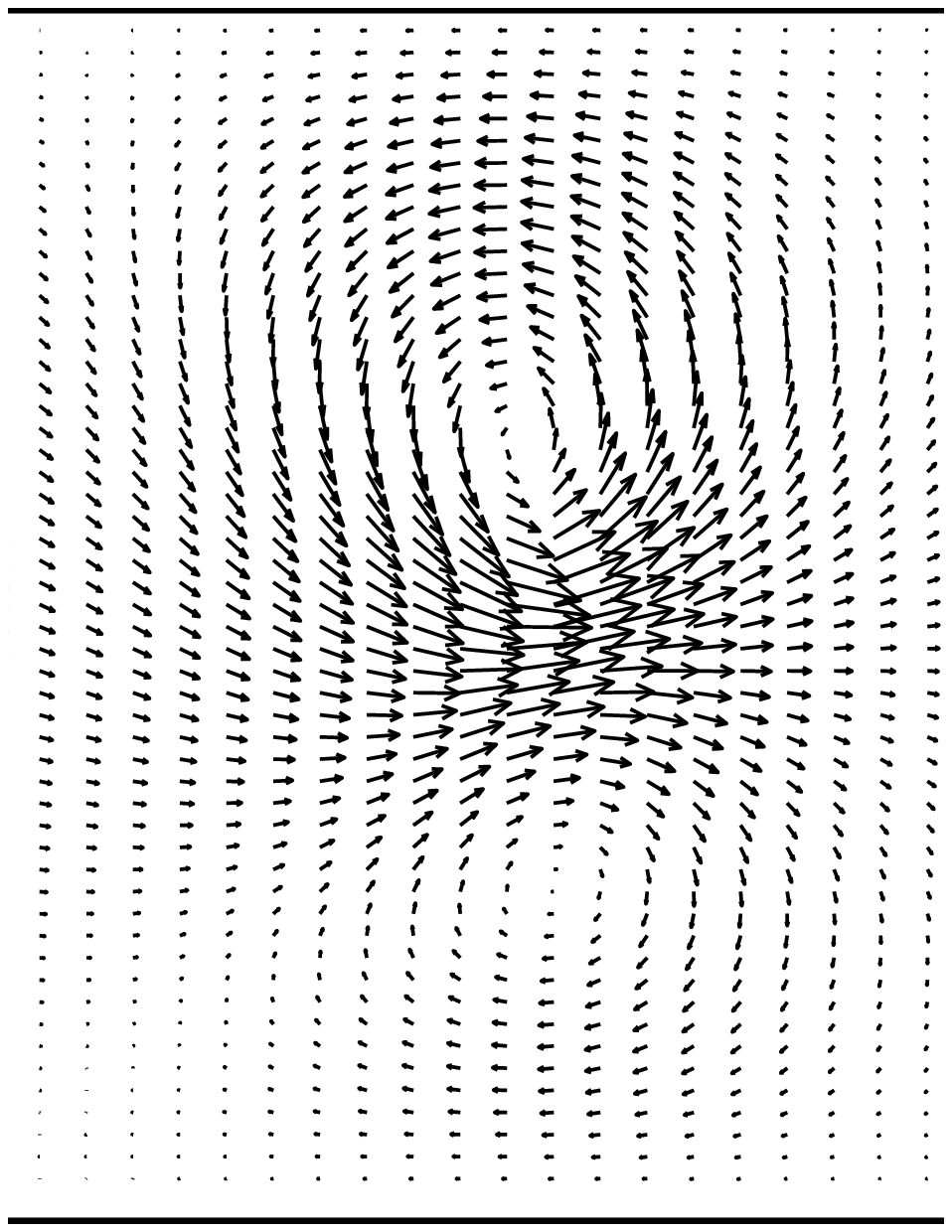}}
\centerline{(b) \hskip 1.6in (c)}
\vskip 0.2true cm
\caption{(a) Director and velocity field about the (b) $+1/2$ and (c)
  $-1/2$ defects for asymmetric surface anchoring.
The director tilt is $\theta_{p}(x=0)=-10\deg$ at the top and
$\theta_{p}(x=L_x)=+60\deg$ at the bottom surface.
The qualitative differences between the flow-fields of the two defects 
are the same as for the symmetric case in Fig. \ref{fig_velocityfield}.}
\label{fig_asym}
\end{figure}

Changing $A_0$ in the free energy (\ref{eqBulkFree}) 
does not affect the homogeneously aligned bulk H
and V domains, only the defect structure. The larger $A_0$, the larger the
energy cost of any deviation from the magnitude of order corresponding to
the minimum of $f_{bulk}$ in (\ref{eqBulkFree}). Decreasing $A_0$
increases the size of the defects, and as above, the defects
move faster under the relaxational dynamics.
Since the defect core size increases, the magnitude of order
around the core decreases resulting a smaller backflow due to
reorientation and hence $\alpha$ decreases. 
The effect of increasing $L_1$ is similar.
It increases the defect core size, speeds up the defects, and leads 
to a smaller velocity anisotropy.

{\bf Non-symmetric surface tilt:}
The director surface tilt at the top and bottom surfaces does not have
to be symmetric. In this more general case, tilt angles at the
surfaces can be written as a sum of a symmetric and an antisymmetric
contribution: 
\begin{eqnarray}
\theta_{p}(x=0)  &=& -\theta_{p,s}+\theta_{p,a},\\
\theta_{p}(x=L_x) &=& +\theta_{p,s}+\theta_{p,a}. \nonumber
\end{eqnarray}
The dynamical equations for $L_2=L_3=0$
and without flow are invariant under the local rotation of all the molecules
by the same angle. The dynamics for a non-symmetric surface tilt can
therefore be obtained 
from the symmetric case by rotating all the molecules by $\theta_{p,a}$.
Thus even for non-symmetric tilts the two defects move with the
same speed and remain in the center between the two plates.
The full dynamical equations however,  
are not invariant under the rotation of molecules if flow is included
(or if $L_2 \ne 0$, $L_3 \ne 0$).
Thus the defects will move with a different speed.  Moreover they
are no longer constrained by symmetry to lie midway between the two plates.
Typical director and velocity fields for an asymmetric surface tilt are shown
in Fig. \ref{fig_asym}.

\begin{figure}
\narrowtext 
\centerline {\epsfxsize=\columnwidth \epsffile{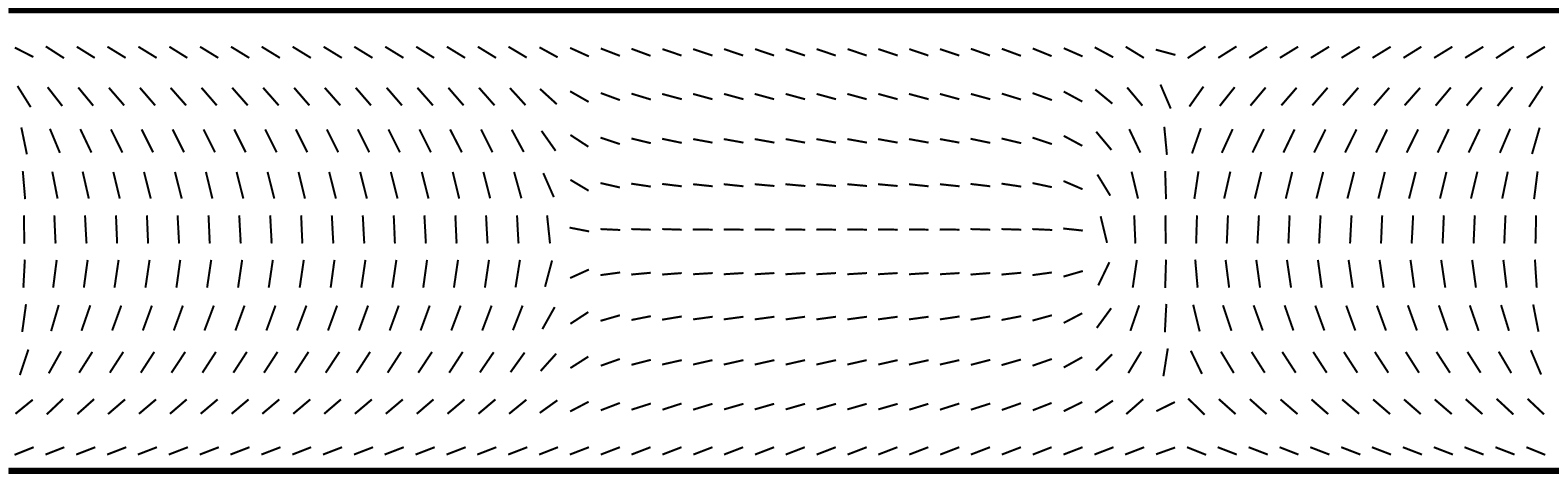} }
\vskip 0.0true cm
\centerline{(a)}
\vskip  0.3true cm
\centerline {\epsfxsize=\columnwidth \epsffile{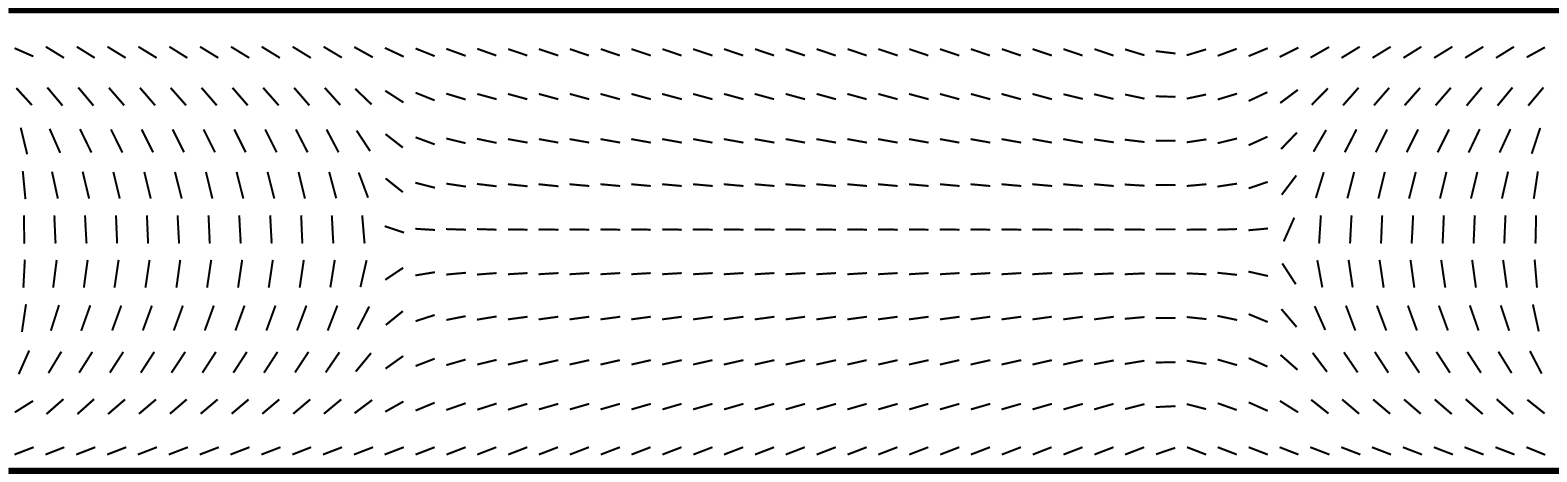} }
\vskip 0.0true cm
\centerline{(b)}
\vskip 0.2true cm
\caption{The influence of surface tilt inhomogeneity:
the surface tilt changes from $\theta_p=+15\deg$ to $\theta_p=-15\deg$
towards the right hand side of the figure. (a) Upon reaching the
border of the two bend domains with opposite surface tilt, 
the defect $+\frac {1} {2}$ merges with the the two $-\frac {1} {2}$
defects located at the surfaces; (b) a $-\frac {1} {2}$ defect is
formed.}
\label{fig_inhom_boundary}
\end{figure}

It is also possible to construct a patterned alignment of liquid crystals
on surfaces \cite{GA97}.
If the surface tilt is not homogeneous then, when the defect arrives
at a region with a different tilt, it assumes the new velocity
corresponding to this tilt. On the boundary of two bend 
domains with opposite surface tilt the defect can even ``change''
topological strength by merging with the two defects located 
at the surface as shown in Fig. \ref{fig_inhom_boundary}.

\section{Growth of a cylindrical domain: the hydrodynamics of a defect ring}

In this section we consider the three dimensional analogue of our system, 
where a liquid crystal is held between parallel plates $\sim \mu m$
apart. A domain nucleated at a point will grow to a cylindrical shape
with its axis perpendicular to the confining plates.  Based on our
experience with the two dimensional system we can discuss how
backflow can affect the growth of a cylindrical H domain in a V environment. 

\begin{figure}
\narrowtext
\centerline{\epsfxsize=3.2in \epsffile{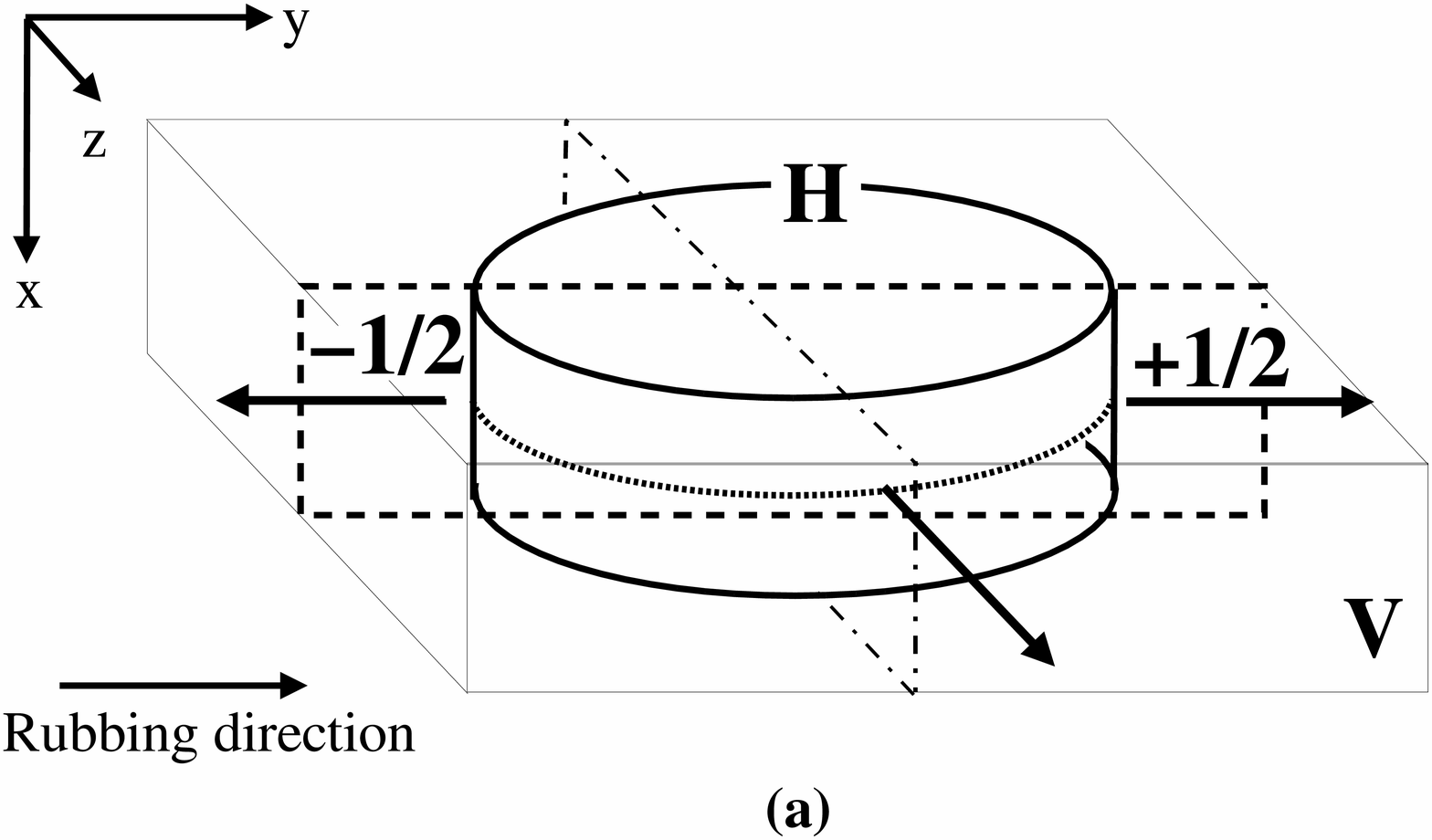}}
\centerline{\epsfxsize=3.2in \epsffile{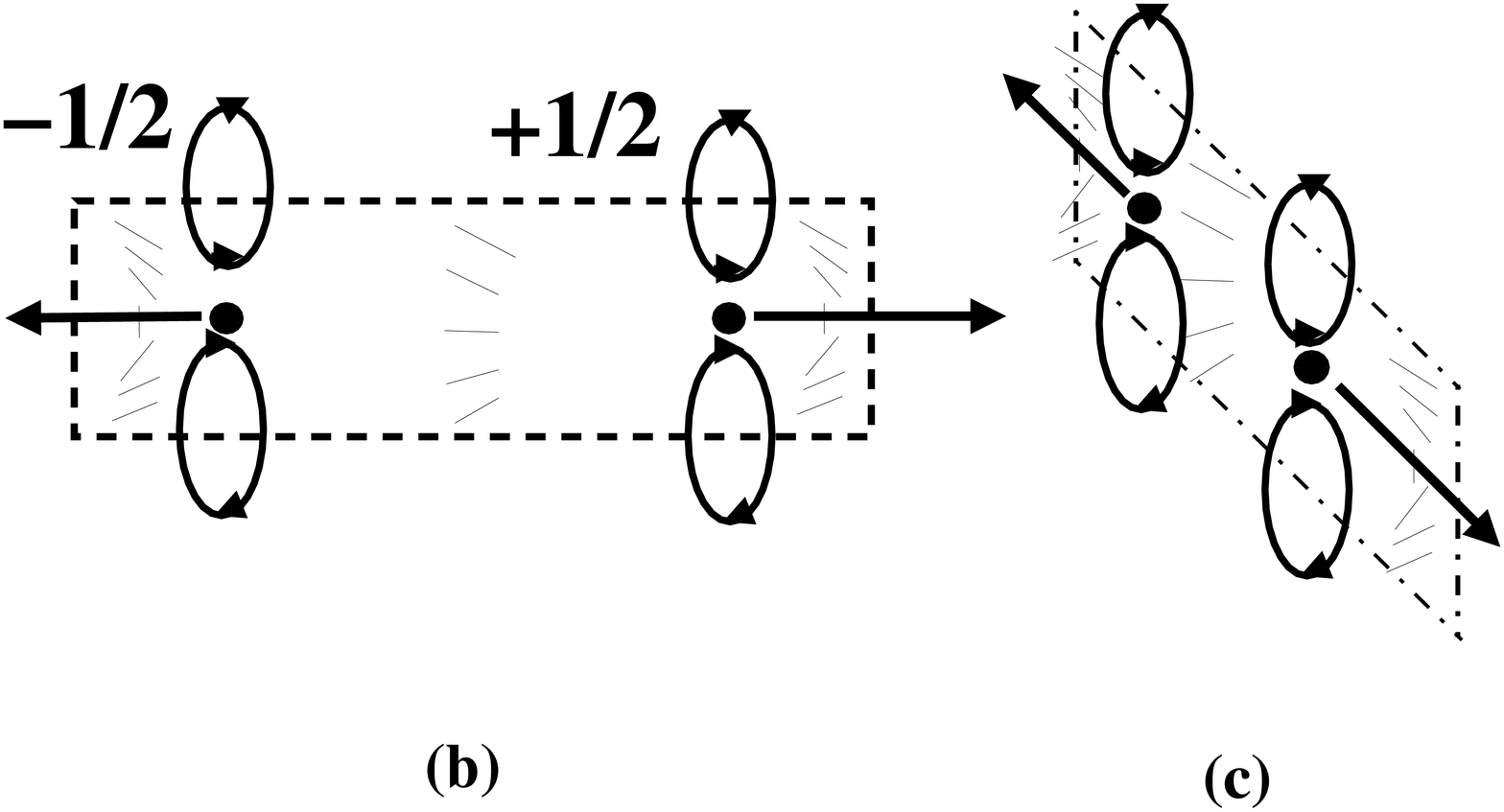}}
\caption{
(a)
Confined between two horizontal surfaces, 
a cylindrical H domain is growing in a V environment. 
There is a defect ring (dotted line) at the domain boundary. 
(b) The cross sections indicated by dashed lines in (a) are shown.
This corresponds to the simulation plane considered in this paper.
The director and the vortices due to the reorientation are in the plane of the cross section.
(c) The cross section indicated by dashed-dotted lines in (a).
The directors are perpendicular to the plane of the cross section. 
The plane of vortices due to the reorientation
is also perpendicular to the cross section. However, for both (b) and (c) 
the vortices due to the deformation stress are in the plane of the cross section and the
flow at the core points in the direction of defect propagation indicated by straight arrows.}
\label{fig_reorient3D}
\end{figure}

At the domain boundary there is a defect ring, as shown
in  Fig. \ref{fig_reorient3D}(a). The defect configuration of a vertical 
cross section through the ring (perpendicular to the plates) changes 
gradually from a $-\frac {1} {2}$ to a $+\frac {1} {2}$ defect.
Although this problem is three-dimensional, a vertical cross
section indicated by dashed lines in  Fig. \ref{fig_reorient3D}(a)
gives a geometry similar to that considered in Section IV.

For simplicity, assume that the H domain is a perfect cylinder.  In
this case the director field of any vertical cross 
section passing through the middle of this
cylindrical domain can be obtained from our two-dimensional simulation
plane by rotating the director field locally by a given angle around the
vertical axis, $x$.

Let us now examine the effect of the backflow. 
${\boldsymbol \sigma}_d$ and the diagonal ${\boldsymbol \sigma}_i$
do not change during the local rotation 
of the tensor order parameter field ${\bf Q}$ around the vertical axis, $x$, 
by the same angle.  The generated velocity vortices always lie in the
plane of the cross sections across the defects ring. These generate
flow which is always in such a direction as to expand the defect ring,
indicated by straight arrows in Figs. \ref{fig_reorient3D}(a-c).

The tensor ${\boldsymbol \sigma}_H$ does however 
change under a local rotation around $x$.
For the cross section which includes a $+\frac {1} {2}$ defect
the resultant backflow points in the
direction of defect motion. For the $-\frac {1} {2}$ defect, 
the flow points opposite to the defect motion, as shown in
Fig. \ref{fig_reorient3D}(b). In both cases the vortices are in the plane of
the cross section.
Fig. \ref{fig_reorient3D}(c) shows the cross section indicated by the
dashed-dotted lines in Fig. \ref{fig_reorient3D}(a). 
Now the directors and the vortices due to
the reorientation are in a plane perpendicular to the cross section.
Thus the total flow field will vary around the domain wall and will
lead to anisotropy in the domain growth.

An experimental setup similar to this was considered
by Acosta {\it et al.} in their investigation of domain growth
and switching in pi-cell liquid crystal devices\cite{AT00}. 
The growth of a horizontal domain in a bend (V) or twisted bend environment 
was studied: such a transition is needed to produce the
operational state of the device.
(The twisted bend configuration has a lower energy than the bend state 
for small surface tilts and low voltage, if the
Frank elastic constant $K_2$ is sufficiently small. 
The twisted bend state is replaced by a bend state for larger ($ \sim 30\deg$) surface tilts.)
A cylindrical bend or twisted bend domain was formed in a H environment
and  the domain wall velocity was measured at four points around the ring,
where its cross section corresponds to a $+\frac {1} {2}$ and a
$-\frac {1} {2}$ defect, and at two points halfway between these.  It
was found that the wall at the $+\frac {1} {2}$ defect moved
substantially faster than at the other three.  It seems very plausible
that the essential physics is captured by our model. 

Further measurements of defect
dynamics in confined geometries have been done very recently \cite{BM02}
and these techniques should allow further testing of the concepts we
present here.

\section{Summary}

In this paper we explored domain growth in nematic liquid crystals.
Defects form at moving domain walls.  We find that a wall
incorporating a $s=+\frac {1} {2}$  defect is substantially speeded up
by backflow effects, whereas a wall containing a $s=-\frac {1} {2}$ 
defect is only slightly affected. This was explained in terms of the
symmetry properties of the different stresses acting on the defects.
These reinforce each other
for the $s=+\frac {1} {2}$ defect while partially cancelling for the
$s=-\frac {1} {2}$ defect.
The influence of different material and geometrical parameters
on the velocity anisotropy was interpreted by comparing the
relative weight of the relaxational dynamics and the backflow.
By generalizing two-dimensional simulation results, 
a qualitative picture was proposed for the role of the backflow
in three dimensions.

Results were obtained using a lattice Boltzmann algorithm to solve
the Beris-Edwards equations of liquid crystal hydrodynamics. 
Working within the framework of a variable tensor order parameter it
was possible to correctly incorporate variations in the magnitude of
order and hence the dynamics of domain walls and their associated
topological defects.

\section{ Acknowledgment }

We would like to thank E.J. Acosta, C.M. Care, 
S. Elston, K. Good, N.J. Mottram, E. Orlandini and
T.J. Sluckin for helpful discussions. We acknowledge the support of
Sharp Laboratories of Europe at Oxford.  C.D. acknowledges funding
from NSF Grant No. 0083286 and G.T. from the EPSRC Grant No. M04426     .

\appendix

\section{A lattice Boltzmann algorithm for liquid crystal hydrodynamics}

We now summarize a lattice Boltzmann algorithm which solves the hydrodynamic
equations of motion of a liquid crystal (\ref{Qevolution}),
(\ref{continuity}), and (\ref{NS}). Lattice Boltzmann algorithms are
defined in terms of a set of continuous variables, usefully termed partial
distribution functions, which move on a lattice in discrete space and time
\cite{CD98}. 

The simplest lattice Boltzmann algorithm, which describes the Navier-Stokes
equations of a simple fluid, is defined in terms of a single set of partial
distribution functions which sum on each site to give the density. For
liquid crystal hydrodynamics this must be supplemented by a second set,
which are tensor variables, and which are related to the tensor order
parameter ${\bf Q}$ \cite{DO00}. 

We define two distribution functions, the scalars $f_i (\vec{x})$ and
the symmetric traceless tensors ${\bf G}_i (\vec{x})$ on each lattice
site $\vec{x}$. Each $f_i$, ${\bf G}_i$ is associated with a lattice
vector ${\vec e}_i$. We choose a nine-velocity model on a square
lattice with velocity vectors ${\vec e}_i=(\pm 1,0),(0,\pm 1), (\pm 1, \pm
1), (0,0)$. Physical variables are defined as moments of the
distribution function
\begin{equation}
\rho=\sum_i f_i, \qquad \rho u_\alpha = \sum_i f_i  e_{i\alpha},
\qquad {\bf Q} = \sum_i {\bf G}_i.
\label{eq1}
\end{equation} 

The distribution functions evolve in a time step $\Delta t$ according to
\begin{eqnarray}
&&f_i({\vec x}+{\vec e}_i \Delta t,t+\Delta t)-f_i({\vec x},t)=\nonumber\\
&&\quad\frac{\Delta t}{2} \left[{\cal C}_{fi}({\vec x},t,\left\{f_i
\right\})+ {\cal C}_{fi}({\vec x}+{\vec e}_i \Delta
t,t+\Delta
t,\left\{f_i^*\right\})\right],\nonumber\\
\label{eq2}\\
&&{\bf G}_i({\vec x}+{\vec e}_i \Delta t,t+\Delta t)-{\bf G}_i({\vec
x},t)= \nonumber\\
&& \quad\frac{\Delta t}{2}\left[ {\cal C}_{{\bf G}i}({\vec
x},t,\left\{{\bf G}_i \right\})+
                {\cal C}_{{\bf G}i}({\vec x}+{\vec e}_i \Delta
                t,t+\Delta t,\left\{{\bf G}_i^*\right\})\right].\nonumber\\
\label{eq3}
\end{eqnarray}
This represents free streaming with velocity ${\vec e}_i$ and a
collision step which allows the distribution to relax towards
equilibrium. 
$f_i^*$ and ${\bf G}_i^*$ are first order approximations to 
$f_i({\vec x}+{\vec e}_i \Delta t,t+\Delta t)$ and ${\bf G}_i({\vec
  x}+{\vec e}_i \Delta t,t+\Delta t)$
respectively. They are obtained by using only the collision operator  
${\cal C}_{fi}({\vec x},t,\left\{f_i\right\})$ on the right of Equation
(\ref{eq2}) and a similar substitution in (\ref{eq3}).
Discretizing in this way, which is similar to a predictor-corrector 
scheme, has the advantages that lattice viscosity terms are eliminated
to second order and that the stability of the scheme is improved.

The collision operators are taken to have the form of a single
relaxation time Boltzmann equation\cite{CD98}, together with a forcing term
\begin{eqnarray}
{\cal C}_{fi}({\vec x},t,\left\{f_i \right\})&=&-\frac{1}{\tau_f}(f_i({\vec x},t)-f_i^{eq}({\vec x},t,\left\{f_i
\right\}))\nonumber\\
& & \quad +p_i({\vec x},t,\left\{f_i \right\}),
\label{eq4}\\
{\cal C}_{{\bf G}i}({\vec x},t,\left\{{\bf G}_i
\right\})&=& -\frac{1}{\tau_{g}}({\bf G}_i({\vec x},t)-{\bf
G}_i^{eq}({\vec x},t,\left\{{\bf G}_i \right\}))\nonumber\\
& & \quad+{\bf M}_i({\vec x},t,\left\{{\bf G}_i \right\}).
\label{eq5}
\end{eqnarray}
The form of the equations of motion and thermodynamic equilibrium
follow from the choice of the moments of the equilibrium distributions
$f^{eq}_i$ and ${\bf G}^{eq}_i$ and the driving terms $p_i$ and
${\bf M}_i$. Full details of the algorithm can be found in \cite{DO00}.

\end{multicols}

\begin{references}
\bibitem{SYM1}
Simulations were performed on a $700 \times 55$
lattice with $A_0=1.0$, $L_1=0.55$, $L_2=L_3=0$, $\Gamma=0.625$,
$\xi=0.59$, $\rho=2.0$, $T=0.5$, $\gamma=3$, $\epsilon_a=41.4$, and
$\epsilon_m=9.8$.  The magnitude of
order $q=\parallel Q\parallel$ is then $0.5$ in the bulk.
Periodic boundary conditions were used in
the $x$ direction and free boundaries were used in the $y$
direction with $\pm 5\deg$ symmetric surface tilt. 
Given suitable pressure, length and time scales these 
parameters can
be mapped to a lattice size $8.8\mu$m$ \times 0.7\mu$m, $L_1=8.73$
pN, $\Gamma=6.25$ Pa$^{-1}$s$^{-1}$. The Frank elastic constants are
$K_{11}=K_{22}=K_{33}=2L_1q^2=4.37$pN and Miesowicz viscosities
$0.02$ Pa s to $0.1$ Pa s. The material parameter values chosen are
close to those of 5CB.
\bibitem{refLx} The distance between the defects is always much larger than 
$L_x$ in the simulations, thus the surface effects dominate the 
defect-defect interaction.
\bibitem{L3=0} If $L_3=0$ and the director is confined to the $xy$ plane,
the Frank elastic free energy density (assuming uniaxiality and constant
magnitude of order) can be written as $f = \frac {\kappa} {2} (\nabla
\theta)^2$, where $\theta$ is the angle of the director to the
horizontal axis and $\kappa=(2L_1+L_2)q^2$.  The dynamics of the
medium, $\partial_t \theta= - \Gamma \delta F / \delta \theta= \Gamma
\kappa \nabla^2 \theta$, are invariant under mirroring on the $x$
axis. Moreover a given $L_1$ and $L_2$ correspond to the same dynamics
as $L_1'=L_1+L_2/2$ and $L_2'=0$.  (Note that there is no speed anisotropy 
in the latter case with the purely relaxational dynamics for the tensor
order parameter.)  For defects this equivalence is only approximate
due to the biaxiality and the change in the magnitude of order in the core.
\end{references}
\end{document}